\documentclass[11pt,draftclsnofoot,peerreviewca,onecolumn,twoside]{IEEEtran}

\usepackage{cite}
\usepackage{amsmath,amssymb,amsfonts}
\usepackage{comment,xcolor}
\newcommand{\z}{\boldsymbol{z}}

\newcommand{\y}{\boldsymbol{y}}
\renewcommand{\v}{\boldsymbol{v}}
\renewcommand{\bf}[1]{\mathbf{#1}}
\usepackage{graphicx,booktabs}
\usepackage{tikz, pgfplots,subcaption}

\usetikzlibrary{external}
\begin{document}
\title{Classification of sums of complex exponentials}
\author{Magdalena Bouza, Andrés Altieri, and Cecilia G. Galarza
\thanks{``This work was supported in part by Agencia I+D+i under Grant PICT 2016-1925 '' }
\thanks{The authors are with the Universidad de Buenos Aires and the CSC-CONICET, Buenos Aires, Argentina (e-mail: cgalarza@csc.conicet.gov.ar). }}

\maketitle
	\begin{abstract}
		Numerous signals in relevant signal processing applications can be modeled as a sum of complex exponentials.
		Each exponential term entails a particular property of the modeled physical system, and it is possible to define families of signals that are associated with the complex exponentials. In this paper, we formulate a classification problem for this guiding principle and we propose a data processing strategy. In particular, we exploit the information obtained from the analytical model by combining it with data-driven learning techniques. As a result, we obtain a classification strategy that is robust under modeling uncertainties and experimental perturbations. To assess the performance of the new scheme, we test it with experimental data obtained from the scattering response of targets illuminated with an impulse radio ultra-wideband radar. 
	\end{abstract}
	

\section{Introduction}
\label{sec:introduction}

Classification problems are of utmost relevance in the signal processing domain~\cite{2009fundamentals}. When using time-domain signals, one assumes that the observed signal belongs to one of a finite number of classes. The actual class is unknown to the observer, and his/her task is to assign one while minimizing the possibility of being mistaken. Nevertheless, the classification process demands an appropriate structure to model the observed signal.

One of the most versatile parametric models used in signal processing is the sum of complex exponentials, which is simple yet ubiquitous. Specialists in very different fields such as Ultra WideBand (UWB) radar systems \cite{Rao2000, LeeSarkarMoonSalazar2013}, time-resolved spectroscopy\cite{Jansson2012},  musical signals\cite{Laroche1993}, and electromagnetic scattering~\cite{Shaw2001,Chen2002} use sums of exponentials to describe their experimental observations. Generally, each term in the sum of exponentials models a characteristic of the observed signal, and a particular combination of complex exponential terms may be associated with a signal type. For example, magnetic resonance time-domain signals show spectral lines modeled as exponentially damped sinusoids. When performing material classification within this setup, the problem calls for estimation of the terms in a sum of complex exponentials. Another application is the use of electromagnetic waves for non-invasive and remote detection and sensing applications.  According to the singularity expansion method (SEM)\cite{Baum}, the transient electromagnetic scattered response of complex targets is dominated by damped sinusoids. These are complex natural resonances (CNR) that only depend on the fundamental properties of the target: material composition, electrical properties, size, and shape. Then the set of CNRs represents the target signature. As such, target identification is performed by appropriate estimation of the CNRs \cite{PoyrazSecmen2013}.

 Both examples above illustrate situations where classification problems were built after  time-domain observations of sums of complex exponentials. In this paper, we will tackle this general problem and propose a classification strategy for it. The celebrated Bayesian decision theory, as well as the maximum likelihood principle, are the usual paths to address the classification problem. Nonetheless, they both rely on a priori information that is not always available to the observer. For example, in \cite{ChenShuley2014}, for the particular case of UWB radar signals,  the authors used a Generalized  Likelihood Ratio Test (GLRT) to solve the problem. However, to pose this solution, they have assumed that the CNRs were known or identifiable with no uncertainty. 
 
 Super-resolution spectral estimation techniques were used in \cite{McCLureQiuCarin1997} for signal identification. In particular,  the authors acquire the signals representing the scattered fields obtained from targets illuminated by UWB signals. Their principal strategy was to exploit the signal model structure that is a superposition of damped exponentials. However, the worked examples did not include experiments where the complex exponentials remained clustered in a small region, which is a problem for the spectral estimation algorithm.  Although several high-resolution spectral estimation techniques are widely available, accurate identification of damped resonances is still a challenge in practice. Then, to solve the classification problem, we require to consider some ambiguity in the signal model. 

Machine learning techniques have become quite popular to solve classification problems. Part of their power lies in their ability to build complex models from a set of observations or training examples. However, the observation structure is decisive in the classification performance. The focus of this work is to explore appropriate data pre-processing strategies for solving the classification problem when observing sums of complex exponentials. In particular, we propose a model-based feature-building procedure that combines the information from the analytical model with data-driven learning techniques. To test our strategy, we will use a UWB radar to perform target classification for its composition. Learning techniques have been used already for target identification using UWB radars \cite{Milani2019}. The results are encouraging, but the approach is solely data-driven, and there is no indication of how this could be generalized to other types of measurements.

In this work, we use the complex natural frequencies associated with the signals to perform classification. While this approach has already been studied in the literature, it has been applied when the natural frequencies were distinct. This is the case of perfectly conducting targets illuminated by radio signals~\cite{PhDAustralia}. 
A different example is presented in \cite{Bannis2014}, where the authors use natural resonances to identify the presence of breast cancer. Also, in \cite{Garzon}, the authors proposed a method for classifying dielectric targets using natural frequencies, where results were validated by  classification of four different target sizes. However, in this case, one of the natural resonances was inversely proportional to the target size.

In this paper, we introduce a processing strategy that proceeds in three steps. The first one solves a spectral estimation problem where the model order and the complex natural frequencies are extracted. This step is performed for a collection of labeled signals, where the label corresponds to the signal class. The second step arranges the estimated natural frequencies to define a feature vector representing the signal. In the third step, we train a classifier using the feature vectors. In this work, we chose to work with a support vector machine. This is a simple nonlinear classification structure that is robust to outliers and has good performance even when the number of training samples is low \cite{burges}. 
We test our framework on numerical examples but also experimental measurements. In particular, we consider the challenging problem of classifying targets based exclusively in their composition. For this, we consider targets of indistinguishable shapes but composed of different materials. Nonetheless, we show that the proposed approach yields superior classification performance compared with the use of the time-domain signals directly.
In Section \ref{sec:natres}, we introduce the signal model and pose the classification problem that we will address. When doing so, we formulate the processing strategy that motivates this paper. We continue in Section \ref{sec:results} by discussing some results to illustrate the performance of the proposed scheme. Finally, in Section \ref{sec:conclu}, we elaborate some concluding remarks.
\section{Classification Problem}\label{sec:natres}

\newtheorem{problem}{Problem}[section]

\newcommand{\C}{\mathbb{C}}
\newcommand{\R}{\mathbb{R}}
\newcommand{\Rset}{\mathcal{R}}

\subsection{Problem Statement}

Consider signals $x(t)$ described by a sum of complex exponentials:
\begin{equation}
    x(t) = \sum_{i=1}^N \alpha_i z_i^t,\quad t=1,2,\ldots, T,
    \label{eq:xmodel}
\end{equation}
where the scalars $z_i\in \C$ are called the complex natural resonances (CNRs) or natural frequencies, and $\alpha_i\in\C$ are the residues. We build a family of signals by considering small variations of the natural frequencies and the residues, i.e., 
\begin{equation}
\mathcal{F}(z_{1}, \cdots, z_{N}) = \Big\{ 
x:\mathbb{N} \to \C: x(t) =\sum_{i=1}^{N} \alpha_{i} z_{i}^t, \,\,  \alpha_{i} \in \mathcal{B} \subset \C, z_i\in Z_{i}\subset \C, 
i=1, \cdots N \Big\}.
\label{eq:familydef}
\end{equation}
We say that the complex subset $Z_i$ is a  resonance region, which characterizes the uncertainty on each $z_i$.  For instance, in a radar application, the family $\mathcal{F}$ is associated with a particular target
. The resonance regions account for the possible deviations that copies of the same target may show, and the set $\mathcal{B}$ where the residues lie may be associated with different look angles for the target. 

Suppose that we have $P$ families that satisfy  \eqref{eq:familydef}, where the $p$-th family is characterized by $N_p$ natural frequencies $z_{1,p}, \cdots, z_{N_p,p}$. To simplify the notation, we denote $\mathcal{F}_p(z_{1,p}, \cdots, z_{N_p,p})$  as $\mathcal{F}_{p}$. 
We assume that, although the number of classes is known, the number of natural frequencies $N_p$ for each class and their resonance regions $Z_{i,p}$ are unknown. 

We define a noisy observation as
\begin{equation}
y(t) = x(t) + w(t), \quad t = 1, \cdots, T,
\label{eq:ydef}
\end{equation}
where $x(t)$ belongs to one of the $\mathcal{F}_{p}$ classes and $w(t)$ is a perturbation signal. For simplicity, let us define the vector
\begin{equation}
\y = [y(1), \ldots, y(T)]\in \C^{T},
\label{eq:yset}
\end{equation}
which corresponds to a noisy observation.
To extract the relevant information which characterizes each class, a labeled set of  observations $\{(\y^{(l)},\rho^{(l)})\}_{l= 1}^L$ is available, where $\rho^{(l)}\in\{1, \cdots P\}$, and $\y^{(l)}$ correspond to noisy observations like in (\ref{eq:yset}).
The problem at hand can be summarized as follows:

\textit{Problem statement:} Given a finite set of labeled noisy observations $\{(\y^{(l)},\rho^{(l)})\}_{l= 1}^L$, where $\rho^{(l)}\in\{1, \cdots P\}$, 
design an algorithm for classifying a new observation $\y$ into one of the $P$ classes.

\subsection{Overview of the proposed solution}

A straightforward approach to solve the classification problem would be to use the time signals directly since they contain implicit information about each class. The downsides to this approach are that time signals can have a large number of samples, leading to an unnecessarily high-dimensional problem. Moreover, they are susceptible to perturbations that can hide the underlying signal model. A more robust approach can be developed by extracting the resonance information given by each $z_i$ in \eqref{eq:xmodel} and classifying a signal directly by using its natural frequencies, which are the base distinctive elements for each class. 
Since these elements are known only through the training samples, a spectral estimation procedure must be employed. There are many spectral estimation algorithms that are specifically tailored to line spectra and are robust against perturbations~\cite{stoica2005spectral}. However, when $N_p$ is unknown, different noisy vectors $\y_p$ from the same class $p$ may lead to different sets of natural frequencies, possibly of different sizes. A key element is to find an appropriate representation that summarizes the spectral information of the observation and at the same time, enables comparing different observations among them. 
With the above problem in mind, we first outline the proposed solution that will be formalized in the next section. The main steps  are as follows:

\begin{enumerate}
\item For each sample $\y^{(l)}$ in the training set, the number and the values of the appropriate natural frequencies are estimated. This involves a spectral estimation procedure that returns a set of natural frequencies. We obtain $L$ complex sets of possibly different sizes.

\item We define a map that relates each complex set found above with an $M$-dimensional complex vector. To obtain $M$, we partition the complex plane into $M$ regions. For that, we consider the union of all the sets of natural frequencies, and we group them into $M$ clusters by using an appropriate clustering algorithm. This problem is not a trivial one, since it entails selecting a proper number of regions and determining which natural frequencies from each observation belong to each region.  Observe that here the clustering procedure only identifies regions of interest that contain a few natural frequencies from at least one sample in the training set, it is not performing signal classification. After determining $M$, we proceed to map each set of natural frequencies obtained for each training signal $\y^{(l)}$ to a vector in $\C^M$. Insofar, the training data is composed of labeled $M$-dimension complex. A dimensionality reduction scheme is also implemented during the training session to retain only the most informative regions.



\item Finally a classifier is trained using the labeled $M$-dimensional vectors obtained before. 

\end{enumerate}

\subsection{Preprocessing}


The essence of our proposal is to preprocess the training data to obtain a suitable structure for the classification procedure. 
As a first step, each signal in the training set is processed individually to obtain its spectral information. For that, we appeal to techniques based on subspace methods as in \cite{Albert2020SpectrumEU}. In the sequel, we outline the procedure for a single signal in the training set. To simplify the notation, we drop the subindex indicating the class of the observation. 

Let $\y$ be a noisy observation of $x(t)=\sum_{i=1}^{N} \alpha_i z_{i}^t$. The problem is to estimate $N$ and the natural frequencies $z_{i}, i=1, \cdots, N$. To cope with noise and uncertainties in $N$, we use optimization techniques that rely on Kronecker's theorem for Hankel operators. Let $\mathbf{H}(x)\in \C^{K\times N}$ represent the Hankel matrix built from the sampled signal $x(1), \cdots, x(T)$, where $T = K+N$. When $N$ is known, 
Kronecker's theorem states that  $\text{rank}(\mathbf{H}(x)) = N$. However, $x(t)$ is only acquired through noisy observations $y(t)$. Then, a suitable denoising procedure is to solve the following optimization problem 
\begin{align}
    &\min_{\mathbf{A}, \tilde{y}}\, \,\, \mathcal{I}_{N}(\mathbf{A}) + \sum_{j=1}^{K+N}|y(j)-\tilde{y}(j)| \\ \nonumber
    & \text{s.t.} \quad \mathbf{A} = \mathbf{H}(\tilde{y}),
\end{align}
where $\mathcal{I}_{N}(\mathbf{A})$ is a threshold function associated with the set $\{\mathbf{A}: \text{rank}(\mathbf{A})\leq N\}$ \cite{Andersson2014}.  Now using $\tilde{y}(t)$ we resort to a high resolution spectral estimation technique, such as ESPRIT to obtain the natural frequencies $z_{i}, i=1, \cdots, N$. 

Since $N$ is unknown, model order selection techniques should be performed first, resulting in an estimation $\widehat{N}$. A sensible technique known as ESTER (ESTimation ERror) \cite{Badeau2004} computes an upper bound on the estimation error obtained with ESPRIT and selects the model order that minimizes such bound. A related approach, known as SAMOS (Subspace-based Automatic Model Order Selection), was introduced in \cite{Papy2007}. However, it has been observed that both techniques have
poor performance in noisy environments. On the other hand, using a hard threshold as in \cite{gavish2} to truncate the  Hankel matrix $\mathbf{H}(y)$ tends to overestimate the model order when the SNR is high. In this work, we have used a combined scheme, as the one presented in \cite{AlbertGalarzaUrucon}, where a constrained optimization problem is posed with the function used in ESTER (or SAMOS). 


The spectral estimation procedure transforms a noisy  observation $\y$ into a set of estimated natural frequencies $\{\hat{z}_y\}_{i=1}^{\widehat{N}}$, where $\widehat{N}$ is the estimated model order. To compare 
two signals, we would like to consider their corresponding sets of natural frequencies. However, natural frequencies are complex numbers that cannot be sorted easily, complicating the task of pairing the natural frequencies to perform the comparison. An alternative, is to map each complex natural frequency to a high dimension vector space and to perform the comparison there.

Let the union of all the sets of natural frequencies obtained for all the training vectors  $\y^{(1)}, \ldots, \y^{(L)}$, be 
$\mathcal{Z} = \{\hat{z}_1^{(1)},\ldots,\hat{z}_{\hat{N}^{(1)}}^{(1)},\ldots, \hat{z}_1^{(L)},\ldots, \hat{z}_{\hat{N}^{(L)}}^{(L)}\}\subset \C$. In general, the natural frequencies $\hat{z}_i^{(j)}$ tend to be concentrated on some regions in the complex plane. Our goal is to identify those regions  to define a finite partition of the complex plane. For that, we propose to apply a clustering procedure on $\mathcal{Z}$.  In particular, we use an agglomerative clustering scheme with complete linkage  and Euclidean distance\cite{hiearchical_cluster}. This is a handy  technique that does not require previous knowledge of the number of clusters, and it does not make assumptions on the statistical properties of the data, such as Gaussian distribution or equal standard deviation among clusters. 

The procedure starts by considering that each natural frequency is in a cluster of its own. 
There are $Q=\sum_{i=1}^{L}\hat{N}^{(i)}$ initial clusters. 
The distance between two clusters $c_i$ and $c_j$ is \begin{equation}
d(c_i,c_j) = \underset{{\hat{z}_l} \in c_i,\ {\hat{z}_m}\in c_j}{\max} |\hat{z}_{l} - \hat{z}_{m}|. 
\end{equation}
At each step, the clusters with minimum distance are merged into a single cluster, and the cluster set is updated. Let $\mathcal{C}_{k-1}$ be the cluster set at iteration $k-1$. Then, $\mathcal{C}_{k}$ is obtained as:
\begin{equation}
\mathcal{C}_k = \mathcal{C}_{k-1} \backslash c_1^k \backslash \  c_2^k \cup c^{new},
\end{equation}
where $c_1^k,c_2^k = \underset{c_i,c_j}{\arg\min}\ d(c_i, c_j)$ and $c^{new} = c_1^k \cup c_2^k$.
The process is repeated until $d(c_i^k , c_j^k) > d_{th}$, for all $i,j$. After finishing the clustering process, there are $M\leq Q$ clusters that define a partition of the complex plane. Now, we proceed to map the set of estimated natural frequencies $\widehat{z}$ to a vector in $\tilde{v}\in \C^M$. For that, we compute the $k$-th element of $\tilde{v}$ by averaging the elements of $\widehat{z}$ found in the $k$-th region of the $M$-partition of $\C$. If $\widehat{z}$ does not have any resonance in the $k$-th region, then $\tilde{\v}(k)=0$. Notice that when applying this operation to the training data, we map sets of different cardinality, $\widehat{\z}^{(1)}, \cdots, \widehat{\z}^{(L)} $, into vectors that lie on the same space $\C^M$. The vector $\tilde{\v}\in \C^M$  is the feature vector. 


\subsection{Classifier}




After preprocessing the training data, we obtain a set of features $\{\tilde{\v}^{(l)}\}_{l=1}^L$. To reduce the problem dimensionality, we select the most explicative features for the classification problem. We propose to use a univariate test, namely the Kruskal-Wallis test, on each feature to quantify its relevance. This is a non-parametric rank test, which tests, for each feature, the null hypothesis that all $P$ classes have a common mean \cite{Kruskal}. 
Higher values of the associated statistic build greater evidence against the null hypothesis, which implies more informative features. We keep the $c$\% more representative features, where the value of $c$ is a hyperparameter of the model.

\begin{figure}
    \centering
    \includegraphics[width=\linewidth]{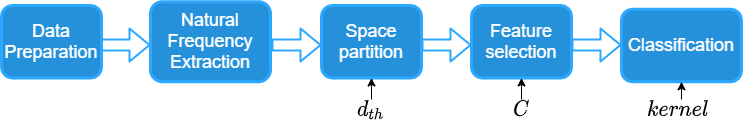}
    \caption{Preprocessing procedure and classification strategy.}
    \label{fig:block_diag}
\end{figure}

After the feature selection, the training set results $\{(\v^{(l)},\rho^{(l))})\}_{l=1}^L$, 
where $\v^{(l)}\in \C^c$ is a feature vector and $\rho^{(l))}$ its corresponding label. We build a $P$-ary classifier using this training data. For that, we work with Support Vector Machines (SVM). The advantages of these classifiers are twofold: on one hand, they allow for nonlinear classification when we use nonlinear kernels; on the other hand, they perform well under small training set size, even when the number of samples is lower than the number of features. In addition, SVMs are robust to outliers and present good generalization performance \cite{burges}. To use SVMs with real inputs, we concatenate real and imaginary parts of $\v^{(l)}$ in the training set.
 Fig.  \ref{fig:block_diag} shows a block diagram for the classification strategy already described. Each block has some hyperparameters that are selected for optimal performance during the training process.

\section{Results}
\label{sec:results}

In this section we present the performance of the proposed classification strategy. We start by analyzing synthetic data inspired in real data. Afterwards, we analyze the results of the classification strategy applied to experimental measurements. In both cases, 
we compare the results obtained with the proposed method against the more traditional approach of using the time-domain signals $\y$ as input to the classifiers. For this, we simply train the SVM using the training set $\{(\y^{(l)}, \rho^{(l)})\}_{l=1}^L$. Since we were working with complex signals, the input to this SVM was  $[Re\{\y^{(l)}\}, Im\{\y^{(l)}\}]$, where $Re\{\y^{(l)}\}$ and $Im\{\y^{(l)}\}$ are the real and imaginary parts of $\y^{(l)}$. The SVM input is 2$T$-long. 

In the sequel, we shall refer to the proposed classification strategy based on natural frequencies as NF and the one processing the  time-domain signals directly as TD. To characterize performance of each classification strategy we consider two metrics computed over a set of test samples of size $S$ $\{(\y^{(s)}, \rho^{(s)}), \rho^{(s)}=1,\ldots,P\}_{s=1}^S$. When considering the $s$-th test sample, we say that $\rho^{(s)}$ is its label or the actual class of the sample, and  $\widehat{\rho^{(s)}}$ the predicted one. 
One error metric is the error rate, which is the ratio of incorrectly classified instances to the total number of test samples
\[
\varepsilon =  \frac{\sum_{s=1}^{S} \boldsymbol{1}\{\widehat{\rho^{(s)}} \neq \rho^{(s)}\}}{S}
\]
A second metric is the confusion matrix, whose $i,j$-element represents the percentage of instances belonging to the $i$-the class that were misclassified as belonging to class $j$
\[
\mathbf{E}_{ij} = \frac{\sum_{s=1}^{S} \boldsymbol{1}\{\widehat{\rho^{(s)}}=j , \rho^{(s)}=i\}}{\sum_{s=1}^{S} \boldsymbol{1}\{\rho^{(s)}=i\}}100\%.
\]

\subsection{Synthetic data}

\begin{figure}
    \centering
\begin{tikzpicture}[scale=1]

\definecolor{color0}{rgb}{0.886274509803922,0.290196078431373,0.2}
\definecolor{color1}{rgb}{0.203921568627451,0.541176470588235,0.741176470588235}

\begin{axis}[
axis background/.style={fill=white},
axis line style={white!33.3333333333333!black},
legend cell align={left},
legend style={fill opacity=0.8, draw opacity=1, text opacity=1, at={(0.09,0.5)}, anchor=west, draw=white, fill=white},
tick align=outside,
tick pos=left,
x grid style={white!15!black!20},
xmajorgrids,
xmin=-0.251147342807635, xmax=1.02333852841536,
xtick style={color=white!33.3333333333333!black},
y grid style={white!15!black!20},
ymajorgrids,
ymin=-1.01710627548373, ymax=1.03255691095026,
ytick style={color=white!33.3333333333333!black}
]
\addplot [only marks, mark=*, mark size = 4pt, draw=color0, fill=color0, colormap/viridis]
table{%
0.127542804062867 -0.907514206336906
0.436288577359366 -0.820539473968061
0.931313459174748 -0.164215690349439
0.965407352450679 0.023594047489235
0.573596997319457 0.789488536747053
-0.192829156669281 0.939390402475987
};
\addlegendentry{$\z_1$}
\addplot [only marks, mark=triangle*,  mark size = 4pt, draw=color1, fill=color1, colormap/viridis]
table{%
0.126564004736905 -0.923939767009455
0.436942160872068 -0.88021427534135
0.947710621403599 -0.16710695245744
0.925803711090954 0.0226261553423438
0.526975103646693 0.781272720044543
-0.193216166842953 0.90901059616737
};
\addlegendentry{$\z_2$}
\end{axis}

\end{tikzpicture}
    \caption{Natural frequencies defining $\mathcal{F}_1$ and $\mathcal{F}_2$.}
    \label{fig:natfreq_synthetic}
\end{figure}
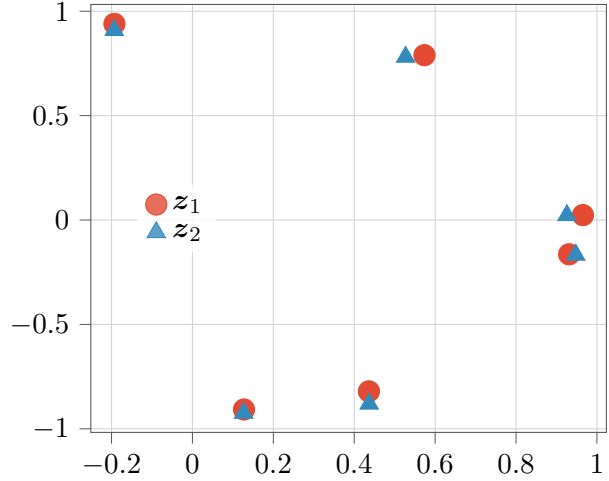

In this subsection we study the performance of the proposed classification approach when applied to synthetic data. The noiseless signals followed the model described in (\ref{eq:xmodel}). For each class of signals as in (\ref{eq:familydef}), we simulated several elements that were used for training and later for testing.  
In particular, we defined two families $\mathcal{F}_p$, $p=1,2$ where the center of each resonance region $Z_{i,p}$ are the nominal frequencies $\z_1 = [ 0.1275 - 0.9075j,  0.44 - 0.16j,  0.97 + 0.02j,
        0.57  + 0.79j, -0.19 + 0.94j]$, and $\z_2 = [ 0.13   - 0.92j,  0.44 - 0.88j, 0.95 - 0.17j,  0.93 + 0.02j,
        0.53 + 0.78j, -0.19+0.91j ]$ as it is shown in Fig \ref{fig:natfreq_synthetic}. We set $T=180$.

\begin{figure}
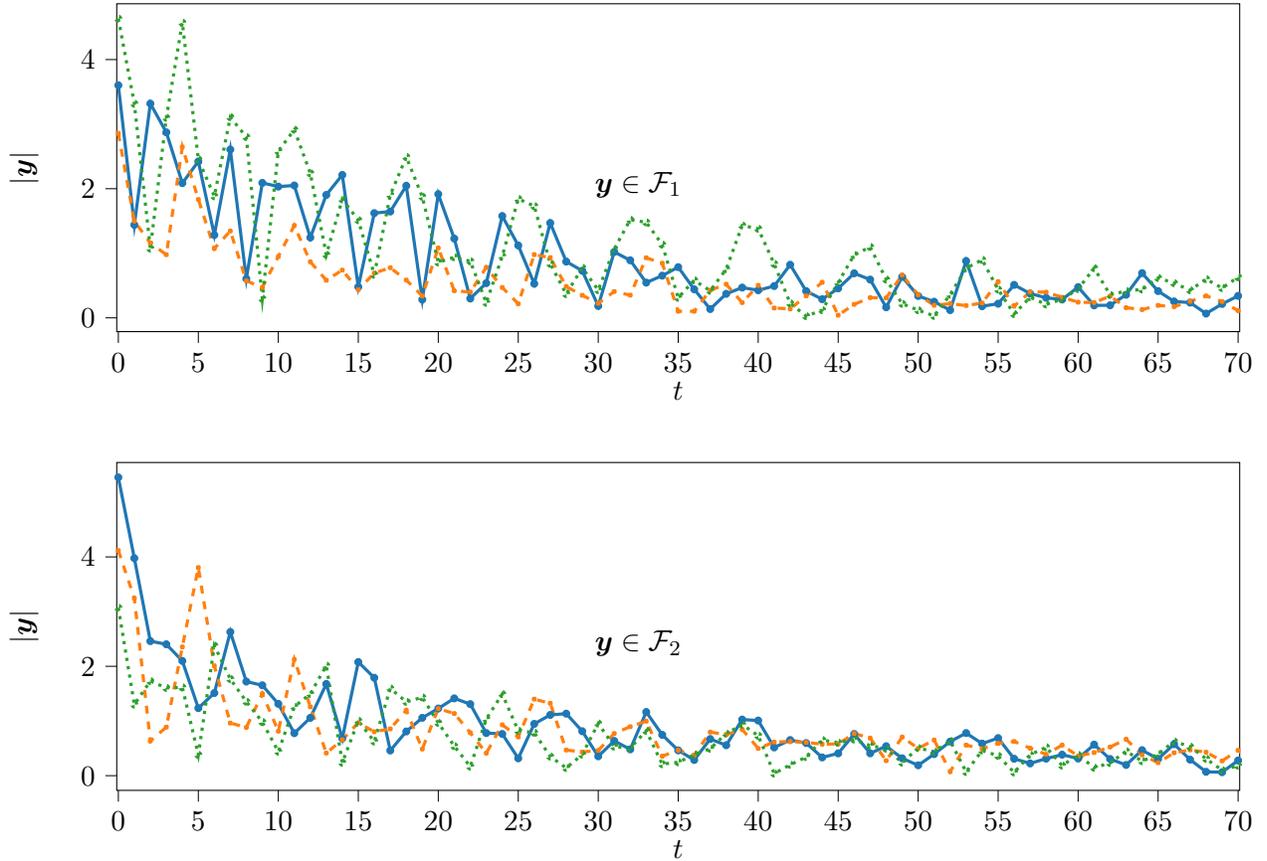

    \centering
    \begin{minipage}{\linewidth}
    \include{x_sinthetic_C1}
    \end{minipage}
    \begin{minipage}{\linewidth}
    \include{x_sinthetic_C2}
    \end{minipage}

    \caption{Training samples using SNR=10dB, $\sigma_\alpha=1$ and the nominal natural frequencies for each class. }
    \label{fig:time_synthetic}
\end{figure}

The corresponding residues for each element of a family were obtained by sampling $\alpha_i$ from a complex Gaussian distribution with mean 0.5 and standard deviation $\sigma_\alpha$. Each observation $\y$ is obtained by adding white noise sampled from a zero-mean circularly symmetric complex normal distribution with variance $\sigma_w^2$.  By varying $\sigma_w^2$, we change the measured signal-to-noise ratio (SNR). We simulated three  scenarios:
\begin{enumerate}
    \item We use the nominal natural frequencies and fix the uncertainty on the residues by considering a fixed value of $\sigma_\alpha^2$.  In particular, we have used 
    $\sigma_\alpha^2=1$. We vary $\sigma_w^2$ to achieve different values for SNR.
    
    \item Once more, we use the nominal natural frequencies, but now we select $\sigma_w^2$ to fix the value of the SNR=10dB. We vary the sets $B_p$ by sampling from distributions with increasing variance $\sigma^2_\alpha$. 
    
    \item In this scenario, we analyze the uncertainties on the natural frequencies. For that, we fix $\sigma_w^2$ and $\sigma_\alpha^2$,  and add perturbations to the natural resonances, which were generated from a complex circularly symmetric normal distribution with zero-mean and variance $\sigma_z^2$. We have used $\sigma_\alpha=1$, and  $\sigma_w$ was selected so SNR = 10dB. 
\end{enumerate}
For each scenario and each value of ($\sigma_\alpha,\ \text{SNR},\sigma_z$), we created three sets of signals for each family for training, while considering 1000 different sets for validation. 
Fig. \ref{fig:time_synthetic} shows $|\y^{(k)}|$ when SNR=10dB, $\sigma_\alpha^2=1$ and the natural frequencies $\z_1$ and $\z_2$ are unperturbed. We notice that variations among the same family are comparable to variations between families. 

\begin{figure}[!ht]
    \centering
\begin{tikzpicture}

\definecolor{color0}{rgb}{0.572549019607843,0.584313725490196,0.568627450980392}

\begin{axis}[
tick pos=left,
xmin=-0.3, xmax=1.1,
ymin=-1.1, ymax=1.1,
legend entries = {{},{$\mathcal{Z}$},{$\z_1$},{$\z_2$}}
]
\addplot graphics [includegraphics cmd=\pgfimage,xmin=-0.3, xmax=1.098, ymin=-1.1, ymax=1.098]
{tikzplots/Voronoi_sinperturbacion-004.png};

\draw[black] (axis cs: -.3, -1.1) -- (axis cs: 1.098, -1.1)
(axis cs: -0.3, -0.94) -- (axis cs: .984, -.94)
(axis cs: -0.3, -0.78) -- (axis cs:0.06, -0.78)
(axis cs: 1.066, -0.78) -- (axis cs: 1.098, -0.78)
(axis cs: 0.858, -0.62) -- (axis cs: 1.098, -0.62)
(axis cs: 0.642, -0.46) -- (axis cs: 1.098, -0.46)
(axis cs: 0.422, -0.29) -- (axis cs:1.098, -0.299)
(axis cs: 0.204, -0.139) -- (axis cs:1.098, -0.139999999999999)
(axis cs:-0.3 , 0.02) -- (axis cs: -0.0219999999999997, 0.0200000000000009)
(axis cs: 1.098, 0.02) --(axis cs: 0.936, 0.02)
(axis cs:-0.3, 0.18) -- (axis cs: 0.0160000000000003, 0.180000000000001)
(axis cs: -0.3, 0.340) -- (axis cs: 0.0540000000000003, 0.340000000000001)
(axis cs: -0.3, 0.5) -- (axis cs: 0.092, 0.5)
(axis cs: 0.692, 0.5) -- (axis cs:0.864, 0.5)
(axis cs: -0.3, 0.660) -- (axis cs: 0.132, 0.66)
(axis cs: 0.61, 0.660) -- (axis cs: 1.098, 0.66)
(axis cs: -0.3, 0.82) -- (axis cs: 0.17, 0.82)
(axis cs: 0.528,0.82) -- (axis cs: 1.09,0.82)
(axis cs: -0.3,0.98) -- (axis cs: 0.208, 0.98)
(axis cs: 0.446, 0.98) -- (axis cs: 1.098, 0.98);


\addplot [draw=color0, fill=color0, mark=*, only marks, opacity=0.8]
table{%
x  y
-0.19433498448605 0.941444335169592
0.44825500697243 -0.825096760712637
0.973718735245065 0.0207697167002751
0.930083483622661 -0.16636757305179
-0.17529512344357 0.941045068615064
0.562093074638107 0.786528204386084
0.962586279369747 0.0259971512094501
0.576664290788415 0.794365900557906
0.437632929918436 -0.880757305655239
0.533193204499267 0.771698116564237
0.913365468723584 0.0361189069705187
0.948346478096099 -0.17260372098744
0.438265253758053 -0.880257707922106
0.938657354851953 0.0947333970176398
0.439474483570612 -0.879303459913136
0.960898025247665 -0.146142502486422
};
\addplot [draw=black, fill=black, mark=x, only marks]
table{%
x  y
0.127542804062867 -0.907514206336906
0.436288577359366 -0.82053947396806
0.931313459174748 -0.164215690349439
0.965407352450679 0.023594047489235
0.573596997319457 0.789488536747053
-0.192829156669281 0.939390402475987
};
\addplot [draw=black, fill=black, mark options={rotate=180}, mark=triangle*, only marks]
table{%
x  y
0.126564004736905 -0.923939767009455
0.436942160872068 -0.88021427534135
0.947710621403599 -0.16710695245744
0.925803711090954 0.0226261553423438
0.526975103646694 0.781272720044543
-0.193216166842953 0.90901059616737
};
\end{axis}

\end{tikzpicture}
    \caption{Partition of the complex plane when using SNR=10dB, $\sigma_\alpha^2=1$ and nominal  natural frequencies. Striped regions are those kept after feature selection process. Grey dots are the estimated natural frequencies, while black crosses and triangles are the the values of $\mathcal{F}_1$ and $\mathcal{F}_2$ respectively.}
    \label{fig:Regions1}
\end{figure}
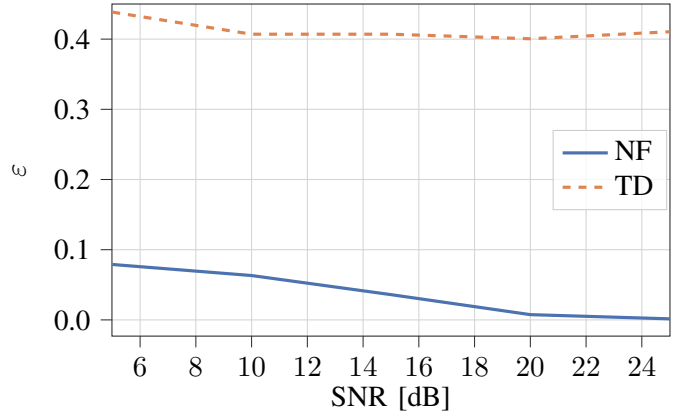

First, we analyze the preprocessing procedure for given values for $\sigma_\alpha^2$, $\sigma_w^2$ and using the nominal natural frequencies. Fig. \ref{fig:Regions1} shows the estimated and true resonances in the training set along with the partition in $M$ regions obtained for the complex plane. In this case, the termination threshold for the clustering algorithm was $d_{th} = 0.03$, which was the optimal value found via cross-validation. Using this threshold, we obtained a partition of size $M=9$. After performing the feature selection step, we kept the regions associated with the most representative features, which are shown as striped regions in Fig. \ref{fig:Regions1}. We managed to reduce almost by half the number of training features. This shows a significant reduction in the number of inputs to the classifier: while we only need five regions (10 features)  to train the SVM under the NF strategy, the corresponding time signal consists of $T=180$ points (360 features).

\begin{table}
    \centering
    \begin{tabular}{|c|c|c|}\hline
   Strategy  & $\varepsilon$ & $\mathbf{E}$  \\\hline
         NF & 0.073 & \begin{tabular}{c c c}
         \vspace{-2pt} & \vspace{-2pt} & \vspace{-2pt}\\
         & $\mathcal{F}_1$ & $\mathcal{F}_2$\\
         $\mathcal{F}_1$ & 90.1\% & 9.9\% \\
         $\mathcal{F}_2$ &4.6\% & 95.4\%\\
         \vspace{-2pt} & \vspace{-2pt} & \vspace{-2pt}\\
         \end{tabular} \\ \specialrule{1.2pt}{1pt}{1pt}
         
     TD & 0.331 & \begin{tabular}{c c c}
     \vspace{-2pt} & \vspace{-2pt} & \vspace{-2pt}\\
      & $\mathcal{F}_1$ & $\mathcal{F}_2$\\
         $\mathcal{F}_1$ & 60.4\% & 39.6\% \\
         $\mathcal{F}_2$ &26.5\% & 73.5\%\\
         \vspace{-2pt} & \vspace{-2pt} & \vspace{-2pt}\\
     \end{tabular} \\\hline
\end{tabular}
    \caption{Comparison of performance  on synthetic data when SNR = 10dB, $\sigma_\alpha=1$ and nominal natural frequencies. }
    \label{tab:err_synthetic}
\end{table}

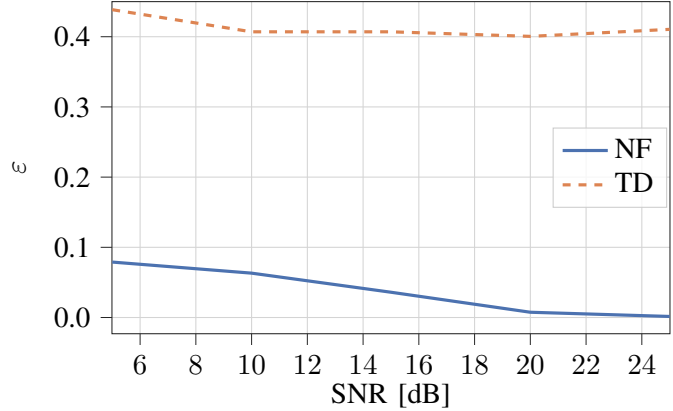
\begin{figure}[!ht]
    \centering
\begin{tikzpicture}

\definecolor{color0}{rgb}{0.917647058823529,0.917647058823529,0.949019607843137}
\definecolor{color1}{rgb}{0.298039215686275,0.447058823529412,0.690196078431373}
\definecolor{color2}{rgb}{0.866666666666667,0.517647058823529,0.32156862745098}

\begin{axis}[
axis background/.style={fill=white},
axis line style={white!15!black},
legend cell align={left},
legend style={fill opacity=0.8, draw opacity=1, text opacity=1, at={(0.995,0.5)}, anchor=east, draw=white!80!black, fill=white},
tick align=outside,
width=9cm,
height=6cm,
x grid style={white!15!black!20},
tick pos=left,
xlabel={SNR [dB]},
xmajorgrids,
xmin=4.99, xmax=25.01,
xtick style={color=white!15!black},
y grid style={white!15!black!20},
ylabel={$\varepsilon$},
ymajorgrids,
ymin=-0.0231, ymax=0.45,
ytick style={color=white!15!black},
ytick={-0.1,0,0.1,0.2,0.3,0.4},
yticklabels={−0.1,0.0,0.1,0.2,0.3,0.4,0.5}
]
\addplot [very thick, color1]
table {%
5 0.07898009950248752
10 0.06319514661274017
15 0.03600000000000003
20 0.007499999999999951
25 0.0014999999999999458
};
\addlegendentry{NF}
\addplot [very thick, color2, dashed]
table {%
5 0.4385
10 0.40700000000000003
15 0.40700000000000003
20 0.40049999999999997
25 0.4105
};
\addlegendentry{TD}
\addlegendentry{Time signal}
\end{axis}

\end{tikzpicture}
    \caption{Classification error for scenario 1).}
    \label{fig:Err_synthetic1}
\end{figure}

Table \ref{tab:err_synthetic} shows the error and confusion matrix of both strategies for the same scenario. As anticipated, the time-domain approach has poor performance since it cannot differentiate between variations among the same family from those arising among classes. 
Figure \ref{fig:Err_synthetic1} shows the error obtained for scenario 1) for different values of SNR [dB], for both the NF and TD strategies. We observe that for all tested SNR levels the classification performance based on natural frequencies is significantly better than the one based on time-domain signals. Additionally, we see a significant improvement as the SNR increases  for the NF classifier, which does not happen under the TD approach. The TD strategy relies heavily on the values of the residues associated with each natural frequency in model \eqref{eq:familydef}. Variations in residues lead to sensible disparities of the time-domain signals within the same family, as we can see in Fig. \ref{fig:time_synthetic}. Consequently, the classifier has poor performance when separating families. Therefore an increase in the SNR does very little in decreasing the error. However, by extracting the natural frequencies, the NF strategy becomes independent of the residues. Moreover, an increase in the SNR results in a more accurate estimation of the resonances, improving the error of the classifier.

\begin{figure}[!ht]
    \centering
\begin{tikzpicture}

\definecolor{color0}{rgb}{0.917647058823529,0.917647058823529,0.949019607843137}
\definecolor{color1}{rgb}{0.298039215686275,0.447058823529412,0.690196078431373}
\definecolor{color2}{rgb}{0.866666666666667,0.517647058823529,0.32156862745098}

\begin{axis}[
axis background/.style={fill=white},
axis line style={white!15!black},
legend cell align={left},
legend style={fill opacity=0.8, draw opacity=1, text opacity=1,  at={(0.005,0.97)}, anchor=north west, draw=white!80!black, fill=white},
tick align=outside,
width=9cm,
height=5cm,
x grid style={white!15!black!20},
tick pos=left,
xlabel={$\sigma_\alpha$},
xmajorgrids,
xmin=-1.302, xmax=1.01,
xtick style={color=white!15!black},
xtick={-1.30103, -1.     , -0.30103,  0.     ,  0.69897, 1},
xticklabels={0.05,0.1,0.5,1,5,10},
y grid style={white!15!black!20},
ylabel={$\varepsilon$},
ymajorgrids,
ymin=-0.0231, ymax=0.51,
ytick style={color=white!15!black},
ytick={-0.1,0,0.1,0.2,0.3,0.4,0.5},
yticklabels={−0.1,0.0,0.1,0.2,0.3,0.4,0.5}
]
\addplot [very thick, color1]
table {%
-1.30103 0
-1 0
-0.30103 0.015007503751875984
0 0.06575619625695495
0.69897 0.06968278731149247
1 0.07343750000000004
};
\addlegendentry{NF}
\addplot [very thick, color2, dashed]
table {%
-1.30103 0
-1 0
-0.30103 0.11199999999999999
0 0.4135
0.69897 0.5015000000000001
1 0.501
};
\addlegendentry{TD}

\end{axis}

\end{tikzpicture}
    \caption{Classification error for scenario 2).}
    \label{fig:Err_synthetic2}
\end{figure}
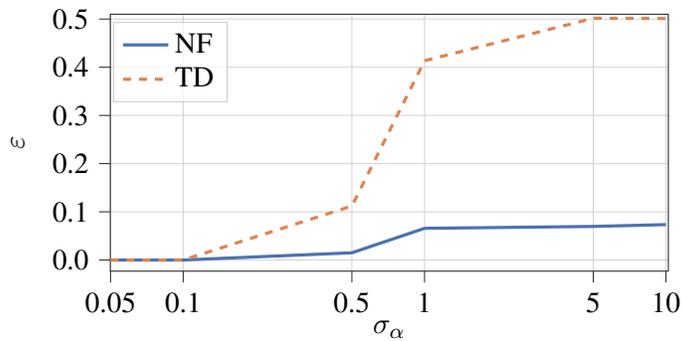

We analyze scenario 2) in Figure \ref{fig:Err_synthetic2}. Here, we have varied the sets $B_p$ by sequentially increasing the variance of the distribution from which the residues are sampled. For sufficiently low variance both classifiers attain zero error. When the sets $B_p$ are small, the residues associated with a natural frequency are quite similar among different realizations of the signals $\y$, making all time-domain signals from the same class very close to each other. By eliminating the variations associated with the residues, even the TD approach achieves no classification error on the test set when SNR=10dB. As $\sigma_\alpha^2$ increases, signals from the same family start to differentiate more from one another, hitting on the classification performance. However, these variations affect much harder the TD strategy compared with the NF strategy. While TD strategy deteriorates its performance, the classification error in the NF strategy remains close to 0.1 even for large values of $\sigma_\alpha^2$.  

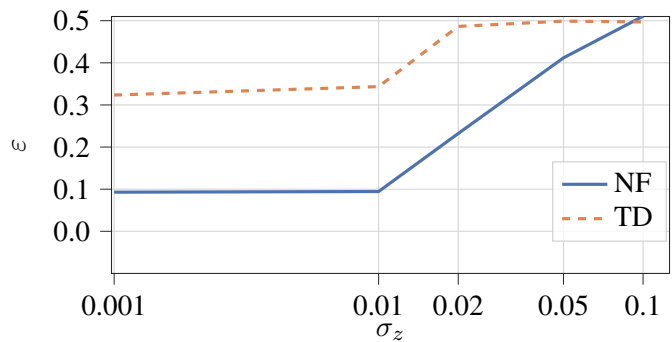
\begin{figure}[!ht]
    \centering
\begin{tikzpicture}

\definecolor{color0}{rgb}{0.917647058823529,0.917647058823529,0.949019607843137}
\definecolor{color1}{rgb}{0.298039215686275,0.447058823529412,0.690196078431373}
\definecolor{color2}{rgb}{0.866666666666667,0.517647058823529,0.32156862745098}

\begin{axis}[
axis background/.style={fill=white},
axis line style={white!15!black},
legend cell align={left},
legend style={fill opacity=0.8, draw opacity=1, text opacity=1, at={(0.995,0.28)}, anchor=east, draw=white!80!black, fill=white},
tick align=outside,
width=9cm,
height=5cm,
x grid style={white!15!black!20},
tick pos=left,
xlabel={$\sigma_z$},
xmajorgrids,
xmin=-3.01, xmax=-0.9,
xtick style={color=white!15!black},
xtick={-3, -2,-1.6989,-1.301,  -1},
xticklabels={0.001,0.01,0.02,0.05,0.1},
y grid style={white!15!black!20},
ylabel={$\varepsilon$},
ymajorgrids,
ymin=-0.1, ymax=0.51,
ytick style={color=white!15!black},
ytick={-0.3,0,0.1,0.2,0.3,0.4,0.5},
yticklabels={−0.3,0.0,0.1,0.2,0.3,0.4,0.5}
]
\addplot [very thick, color1]
table {%
-3 0.09289340101522847
-2 0.09446419502285419
-1.6989 0.2322515212981744
-1.301 0.4118246687054027
-1 0.5106493506493506
};
\addlegendentry{NF}
\addplot [very thick, color2, dashed]
table {%
-3 0.3235
-2 0.3435
-1.6989 0.48650000000000004
-1.301 0.499
-1 0.497
};
\addlegendentry{TD}

\end{axis}

\end{tikzpicture}
    \caption{Classification error for scenario 3).}
    \label{fig:Err_synthetic3}
\end{figure}

Finally,  Fig. \ref{fig:Err_synthetic3} compares the performance of both classifiers for the third scenario.  As the value of $\sigma_z^2$ increases, so does the size of the regions $Z_i$ in \eqref{eq:familydef}.  In this case, the performance of both classifiers drops as perturbations increase. This is expected since natural frequencies from both families were purposely chosen close to each other and, as the resonance regions widens, the overlap between them increases. However, for values of $\sigma_z$ below 0.1 the proposed method presents a significant improvement with respect to the TD classifier. The rapid increase in the error is a result of the overlapping of the resonance regions. By construction, both classes have natural frequencies that are close. The minimum and maximum distances between frequencies are 0.016 and 0.06. This fact explains why we see a  deterioration of the performance when the uncertainty $\sigma_z$ goes above 0.01. At this uncertainty level, some of the natural frequencies of the two classes overlap for certain realizations. Moreover, when $\sigma_z$ is larger than 0.1, the overlapping occurs with a high probability for all the natural resonances. As a result, the scheme loses its prediction ability.

\subsection{Experimental measurements}
In this section we test the performance of the proposed approach when dealing with experimental data. For that we have considered scattering signals from a target illuminated with microwave signals in an experiment similar to \cite{TIM}. The main goal was to classify targets that have the same shape and size, but different material composition.
Each target was illuminated by  an UWB electromagnetic pulse and  the resulting scattering signal was collected by the receiving antenna. When a target is hit by a signal whose spectrum lies in its resonance region, i.e. the wavelength of the incident pulse is on the order of the dimension of the target, the scattering response is the superposition of two signals: one due to direct reflections from the target, called early time, and another one due to a resonance phenomena, called late time \cite{Baum}. This resonance phenomena is determined by the natural  frequencies, which depend solely on the material composition, shape, and size of the target. 
There exists a unique set of natural resonances for each distinct target. According to the Singularity expansion method (SEM) \cite{Baum}, the resonance phenomena, can be modeled as a sum of damped exponentials, as in (\ref{eq:familydef}), thus making our approach suitable  \cite{Chen2014, Bannis2014}. 

For the proposed experiment, we considered a classification problem among plastic bottles containing three different liquids: alcohol (98\%), tap water, and brine, obtained by diluting 35gr of salt in 500ml of water. The liquids were poured in identical containers, with a volume of 500ml. Notice that all the targets have the same shape, weight, and color because all the liquids are transparent. The only differences lie in the composition of the liquids. Traditional approaches based on target image or weight would not work in this case. 

We have used the X4M06 Radar Development kit that is based on the X4 UWB system-on-chip by Novelda~\cite{Novelda}. This platform emits pulses with a bandwidth of 1.5GHz at a central frequency $f_c=8.748$GHz, and receives the scattered response. The captured pass-band response is sampled by the system at an equivalent-time sampling frequency $f_s = 23.328$GS/s and downconverted in the PC to obtain the complex baseband equivalent signal. To improve the signal-to-noise ratio, the platform transmits numerous pulses and then averages the scattered response over multiple transmissions to mitigate random interference and noise.

\begin{figure}
    \centering
    \includegraphics[width = .5\columnwidth]{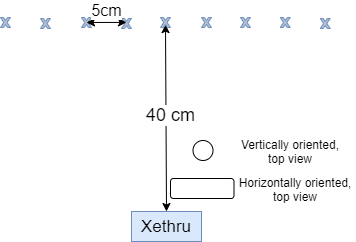}
    \caption{Positioning and orientation of the target.}
    \label{fig:setup}
\end{figure}

\begin{figure}
    \centering
    \includegraphics[width=0.7\columnwidth]{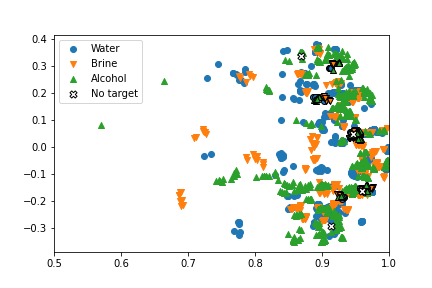}
    \caption{Estimated natural frequencies from the training set. Points with a black edge represent the natural frequencies discarded in the feature selection stage. 
    Light grey points correspond to the resonances obtained when no target is present}
    \label{fig:resonances}
\end{figure}

To diversify the training set, we took several measurements of the scattered response on each target. Targets were located on 9 positions and the scattered signals from each target were collected for each position. For that, we located the targets along a line parallel to the antennas  at a distance of 40cm. On this line, we chose 9 different points separated by 5cm each. Additionally, we oriented each target vertically and horizontally (Fig. \ref{fig:setup}). For each position and orientation, we took 10 measurements  of the scattered response. We denote by $s(t)$ the signal returned by the Xethru, which has a fixed length of $T=1497$ samples. This scattered signal is:
\begin{equation} \label{eq:interference_noise_model}
    s(t) = x(t) + w(t) + d(t), \ \ \ t=1,...,T,
\end{equation}
where $x(t)$ is the true scattered response, $w(t)$ is noise and radio frequency interference from the environment and $d(t)$ is due to antenna cross-coupling and clutter in general. Assuming that the noise and interference in $w(t)$ are stationary in the duration of the transmissions, they are mitigated by the internal averaging process that the transceiver module performs. On the other hand, the disturbance $d(t)$ is persistent and is not mitigated by averaging. Moreover, this disturbance does not have a linear interaction with the scattered response of the target. 
However, one could take an additional measurement with no target present and extract its natural frequencies, which will be associated with reflections from objects of the background. These resonances are also present in the scattering measured from each target, and most of them are discarded in the feature selection stage, as shown in Fig. \ref{fig:resonances}.

\begin{figure}
    \centering
    \input{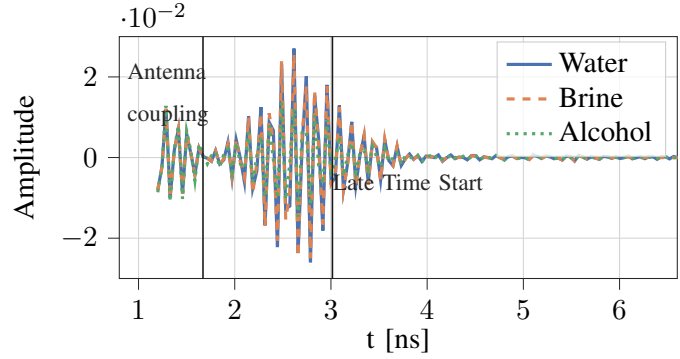}
    \caption{Time domain responses for the different materials. These measurements correspond to the targets located in front of the antennas at a distance of 40 cm, horizontally aligned. }
    \label{fig:time_UWB}
\end{figure}

Fig. \ref{fig:time_UWB} shows the signals received by the Xethru for each target located at 40cm of the antennas and vertically aligned. This figure depicts the antenna coupling signal, which appears to be the same across all measurements, and the beginning of the late time response. Notice that for this specific scenario the time-domain responses of the targets filled with water and brine show little difference, while alcohol appears to be easily distinguishable from the other two targets. However this observation is not valid for all measurement setups since signal amplitudes and shapes occurring from different positions and orientations vary widely for the same target, making the classification even more challenging.

To apply model \eqref{eq:xmodel}, we discarded the early time response, and kept the late time response only. By analysing the different captured signals, we determined that the late time began 7 samples after the first peak of the absolute value of the baseband signal. This value was obtained considering the time needed for the incident pulse to pass through the target \cite{PhDAustralia}. Since we considered 9 positions for the target, with 2 possible orientations and recorded 10 different samples, or measurements, of each target at each location, we had a total of 540 labeled noisy observations. These signals were split into training and a testing sets at random, 70\% for training and 30\% for testing. 





We trained two classifiers, one using the NF approach and another one with the TD approach. For fairness we used the same train/test split for both approaches. For the TD strategy, we used the baseband time-domain signals up to 5ns, concatenating real and imaginary parts as described on synthetic data. The second classifier trained under the NF strategy, used the following hyperparameters  $d_{th}=0.03$, $C=0.95$, and a polynomial kernel of degree 2 for the SVM. 

Figure \ref{fig:resonances} shows the natural frequencies extracted from the training data set. Points with a black edge correspond to features discarded in the feature selection stage of the training algorithm. Additionally, grey points correspond to the natural frequencies found when no target was present, which could be associated with the disturbance signal $d(t)$ in \eqref{eq:interference_noise_model}. Notice that most of those are associated with discarded features.

The reported performance of the NF and TD classifiers was calculated over the test set. Table \ref{tab:err_UWB} summarizes the results for both classifiers. We see that in both cases, alcohol is better classified than the other two substances. However, the overall performance of the ND classifier is two orders of magnitude better than the TD classifier. Moreover, the TD classifier confuses brine with alcohol in 40\% of the test samples, while the ND classifier has a perfect score for the same test samples. 

Additionally, the training features of the NF method were stored in a sparse matrix, since each target has natural frequencies located in a few clusters. By exploiting sparsity, we benefit from more efficient data storage, which is an added advantage of the proposed classification strategy. Also, we observe a reduction in the number of features. After feature selection, only 103 clusters (206 features)  were kept, while the time-domain approach used 117 sample points (234 features).  
\begin{table}[ht]
    \centering
    \begin{tabular}{|c|c|c|}\hline
        \vspace{-3pt} & \vspace{-3pt} & \vspace{-3pt}\\
        Strategy & $\varepsilon$ & $\mathbf{E}$ \\
        \vspace{-2pt} & \vspace{-3pt} & \vspace{-3pt}\\\hline
        NF & 0.02 & 
        \begin{tabular}{cccc}
        \vspace{-2pt} & \vspace{-2pt} & \vspace{-2pt}\\& Water & Alcohol & Brine \\
        Water & 95.2\% & 4.8\% &  0\%\\
        Alcohol & 1.7\% & 98.3\% &  0\% \\
        Brine & 0\% &  0\% & 100\% \\
        \vspace{-2pt} & \vspace{-2pt} & \vspace{-2pt}\\\end{tabular}
 \\\specialrule{1.2pt}{1pt}{1pt}
        TD & 0.34 &  \begin{tabular}{cccc} 
        \vspace{-2pt} & \vspace{-2pt} & \vspace{-2pt}\\
        & Water & Alcohol & Brine \\
        Water & 50\% & 32.3\% & 17.7\%\\
        Alcohol & 0\% & 89.7\% & 10.3\%\\ 
        Brine & 0\% & 40\% & 60\%\\
        \vspace{-2pt} & \vspace{-2pt} & \vspace{-2pt}\\
        \end{tabular}\\\hline
    \end{tabular}
    \caption{Comparison of performance of the classifiers.}
    \label{tab:err_UWB}
\end{table}

\section{Concluding Remarks}
\label{sec:conclu}
We presented a new signal classification strategy when the signals are modeled as a sum of complex exponentials. We did this by combining a model-based analysis with data-based learning techniques. Exploiting the analytical model, we utilized the complex natural frequencies as descriptors of each class. Then, we transformed the natural frequencies into characteristics suitable for the classification problem by using statistical data analysis. This signal processing step resulted in a reduction of the dimensionality of the problem as an added benefit. Finally, machine learning techniques performed the classification process. 

We tested the classification method with synthetic signals as well as a real-world problem, namely the classification between different materials using their the scattered responses when illuminated with UWB pulses. We observed a significant improvement in classification performance for all the tests when we confronted our approach with time-domain classification. Moreover, we showed that the overall procedure is robust to multiple uncertainties that arise in the model. For instance, the procedure proposed in this paper tolerates uncertainties in the model order and variations in the residues associated with each natural resonance. In this way, our classification strategy is a suitable candidate for challenging classification problems where imperfect model information or measurement disturbances may hinder the performance of traditional classification methods.

\bibliographystyle{IEEEtran}
\bibliography{biblio_intro2, biblio}

\end{document}